\newcommand{\mb}[1]{\ifmmode#1\else\mbox{$#1$}\fi}

\newcommand\de{\mb{\delta}}

\newcommand\calM{\mb{{\cal M}}}

\newcommand{\beq}{\begin{equation}}
\newcommand{\eeq}{\end{equation}}
\newcommand{\nn}{\nonumber}
\newcommand{\bea}{\begin{eqnarray}}
\newcommand{\eea}{\end{eqnarray}}

\newcommand{\inprod}[2]{\langle {#1}, {#2} \rangle}

\newcommand{\x}{\mb{\times}}

\newcommand{\Ad}{{\rm Ad}}
\newcommand{\gsim}
{\raise.3ex\hbox{$\;>$\kern-.75em\lower1ex\hbox{$\sim$}$\:$}}
\newcommand{\lsim}
{\raise.3ex\hbox{$\;<$\kern-.75em\lower1ex\hbox{$\sim$}$\:$}}

\newcommand{\ts}{\textstyle}
\newcommand{\half}{{\ts \frac{1}{2}}}
\newcommand{\third}{{\ts \frac{1}{3}}}
\newcommand{\twothird}{{\ts \frac{2}{3}}}

\documentstyle[prl,aps,twocolumn,floats]{revtex}

\begin{document}


\twocolumn[\hsize\textwidth\columnwidth\hsize\csname @twocolumnfalse\endcsname
\title{Gauge Interactions in the Dual Standard Model}
\author{Nathan\  F.\  Lepora}
\address{Department of Applied Mathematics and Theoretical Physics,
Cambridge University, England.}  
\date{\today}
\maketitle

\begin{abstract}
We present a geometric argument for the transformation properties of $SU(5)
\rightarrow S(U(3) \x U(2))$ monopoles under the residual gauge symmetry. This
strongly supports the proposal that monopoles of the 
dual standard model interact via a gauge theory of the standard model symmetry
group, with the monopoles having the same spectrum as the standard model
fermions. \\

\ \\
\end{abstract}
]



The dual standard model has been proposed as a way of unifying both
matter and interaction~\cite{vach}. Monopoles from the Georgi-Glashow 
\bea
SU(5) &\rightarrow& S(U(3)\x U(2)) \nn\\ 
&\ &\hspace{4em}= SU(3)_{\rm C} \x U(2)_{\rm I} \x U(1)_{\rm
Y}/{\bf Z}_6 
\eea
grand unification have precisely the same spectrum as the
observed fermions in the standard model; it is therefore natural to associate
these standard model fermions with such monopoles. In consequence of 
this we have calculated the gauge couplings at monopole
unification~\cite{unif} 
\beq
g_{\rm C}/g_{\rm I}=3,\ \ \ 
g_{\rm Y}/g_{\rm I}=2/\sqrt{15}.
\eeq
Both values are satisfied by the standard model gauge couplings at a scale of
a few GeV. 

In this letter we examine the transformation properties of these monopoles
under the residual $S(U(3)\x U(2))$ symmetry. As such we show
that gauge transformations of the fundamental monopoles are entirely
consistent with the fundamental representation of the standard model symmetry
group. This gives strong support to the proposal that the long range
interaction of these monopoles is via a gauge interaction of $SU(3)_{\rm C}$,
$SU(2)_{\rm I}$ and $U(1)_{\rm Y}$ symmetry groups. 

We shall consider firstly the fundamental monopoles. These are embedded $SU(2)
\rightarrow U(1)$ monopoles, 
\bea
\label{embedding}
SU(5) &\rightarrow& S(U(3) \x U(2)) \nn \\
\cup  \hspace{1em} &\ &\  \cup \\
SU(2)_Q &\rightarrow& U(1)_Q. \nn
\eea
It is clear that these fundamental monopoles have a degeneracy of embeddings.
The purpose of this letter to quantify this degeneracy.  

To quantify the space of embeddings we shall label the embedding of
the fundamental monopoles. For this it will prove useful to split the $su(2)$
algebra into components  
\beq
\label{su2dec}
su(2)_Q = u(1)_Q \oplus \calM_Q.
\eeq
Here $u(1)_Q$ is the Lie algebra of $U(1)_Q$, and $\calM_Q$ is its associated
orthogonal component. The direct sum is with respect to the standard inner
product on $su(5)$, given by $\inprod{X}{Y}={\rm tr}XY$.  

One useful label for the fundamental monopoles is their magnetic charge $Q$
(we will see later that there is another more useful label). The magnetic
charge defines the asymptotic magnetic field of a monopole,
\beq
B^k \sim \frac{\hat{r}^k}{r^2}  {Q}.
\eeq
and is associated with the embedding
\beq
\label{embedQ}
U(1)_{Q} = \exp({\bf R} Q) \subset S(U(3)\x U(2)),
\eeq
normalised by 
\beq
\exp(2\pi g_u  {Q})=1,
\eeq
with $g_u$ the unified $SU(5)$ gauge coupling.
Additionally the embedding in Eq.~(\ref{embedQ}) is associated with the
topology of $SU(5)/S(U(3)\x U(2))$, being a non-trivial element of 
\beq
\pi_1[S(U(3)\x U(2))] = \pi_2[SU(5)/S(U(3)\x U(2))].
\eeq
Following \cite{vach}, we decompose the magnetic charge into colour, weak
isospin and weak hypercharge components
\beq
\label{Q}
{Q}=\frac{1}{g_u} \left( T_{\rm C} + \half T_{\rm I} +
\third T_{\rm Y} \right), 
\eeq
where $T_{\rm C}\in su(3)_{\rm C}$ may be either
\bea
\label{deg1}
T_{\rm C}^r&=& i\ 
{\rm diag}(+\twothird,-\third, -\third,0,0),\\ 
\label{deg2}
T_{\rm C}^g&=& i\ 
{\rm diag}(-\third, +\twothird, -\third,0,0),\\ 
\label{deg3}
T_{\rm C}^b&=& i\ 
{\rm diag}(-\third, -\third, +\twothird,0,0),
\eea
and $T_{\rm I}\in su(2)_{\rm I}$ may be either
\beq 
\label{deg4}
T_{\rm I}^\pm = \pm i\  {\rm diag}(0,0,0, 1, -1),
\eeq
whilst $T_{\rm Y}\in u(1)_{\rm Y}$ may only be
\beq
T_{\rm Y} = i\  {\rm diag}(1, 1, 1, -{\ts \frac{3}{2}}, -{\ts
\frac{3}{2}})  
\eeq
The above degeneracies indicate that the fundamental monopoles form
representations of $SU(3)_{\rm C}$, $SU(2)_{\rm I}$ and $U(1)_{\rm Y}$
with the corresponding dimension. Namely the fundamental
representations. 

The purpose of this letter is to investigate these degeneracies. We interpret
the degeneracies as being due to gauge freedom of the monopole embedding. In
this light we show that the gauge degeneracy of the fundamental monopoles
are consistent with the fundamental representions of the residual symmetry
group $S(U(3)\x U(2))$. 

On the issue of duality, we shall show that the dual of the 
residual symmetry group $SU(3)\x SU(2)\x U(1)$ is also consistent with the
gauge degeneracy of the monopoles.

A rigid (or global) gauge transformations of the fundamental monopole is
defined by an element $h \in S(U(3)\x U(2))$ and transforms the magnetic field
as
\beq
\label{mon-gauge}
B^k \mapsto \Ad(h) B^k = h B^k h^{-1}.
\eeq
Correspondingly the $su(2)$ embedding transforms under
\beq
su(2)_Q \mapsto \Ad(h) su(2)_Q,
\eeq
so that $Q$ transforms appropriately. Hence the components of
Eq.~(\ref{su2dec}) transform as
\bea
u(1)_Q &\mapsto& \Ad(h) u(1)_Q,\\
\calM_Q &\mapsto& \Ad(h) \calM_Q.
\eea

One may see that $Q$ is not a good quantity for examining the action of
$S(U(3)\x U(2))$ on the monopole by considering the action of elements $h\in
U(1)_Q$. These take $u(1)_Q\mapsto u(1)_Q$ identically, whilst acting
non-trivially on elements of $\calM_Q$, taking them to another element of
$\calM_Q$. 

Thus to obtain all of the possible monopole embeddings we must examine
the action of $S(U(3)\x U(2))$ on $\calM_Q$. This may 
be achieved by considering the action on any non-trivial element of
$\calM_Q$. Then the manifold of all equivalent fundamental monopoles under a
rigid gauge transformation is  
\beq
\label{M}
M(\calM_Q) \cong \frac{S(U(3) \x U(2))}{C(\calM_Q)},
\eeq
with the centraliser 
\beq
\!C(\calM_Q)\!=\! \{ h \in S(U(3) \x U(2)): \Ad(h)\calM_Q\!=\!\calM_Q\}
\eeq
representing those transformation that leave $\calM_Q$ invariant. 

We shall calculate $C(\calM_Q)$ by considering its action on a monopole
embedding. In particular consider a magnetic charge 
\beq
Q^{r+}=\frac{1}{g_u} \left( T_{\rm C}^r + \half T_{\rm I}^+ +
\third T_{\rm Y} \right), 
\eeq
having explicit components
$g_u Q_{jk}^{r+}=\de_{j1} \de_{k1} - \de_{j5} \de_{k5}$. The $su(2)$ algebra 
associated with this is generated by $\{g_u Q^{r+}, X^{r+}, Y^{r+}\}$,
where the explicit components are
\bea
X_{ij}^{r+} &=& \de_{j5} \de_{k1} - \de_{j1} \de_{k5},\\
Y_{ij}^{r+} &=& i(\de_{j5} \de_{k1} + \de_{j1} \de_{k5}).
\eea
Then $[X^{r+},Y^{r+}]=2 g_u Q^{r+}$.

To exhibit the group structure we require the generators $T_{\rm C}$,
$T_{\rm Y}$ and $T_{\rm I}$ expressed in a basis normalised to the
topology of $SU(5)/S(U(3)\x U(2))$. To this end we define
\bea
C= {\ts \frac{3}{2}} T_{\rm C}^r, \ \ \ \\
I= T_{\rm I}^+,\ \ \ \\
Y= {\ts \frac{2}{5}} T_{\rm Y},
\eea
such that 
\beq
\Ad(e^{2\pi C})\calM_Q\!=\!\Ad(e^{2\pi I})\calM_Q
\!=\!\Ad(e^{2\pi Y})\calM_Q\!=\! 1.
\eeq
In particular
\beq
{\rm Ad}(e^{\theta_C C}){\rm Ad}(e^{\theta_I I}){\rm Ad}(e^{\theta_Y
Y})\calM_Q\!=\!e^{i(\theta_C+\theta_I+\theta_Y)}\calM_Q\!
\eeq
From this we obtain
\beq
C(\calM_Q) = SU(2)_C\x U(1)_{{\rm Y}-{\rm I}} \x U(1)_{{\rm I}+{\rm Y}-2{\rm
C}}/{\bf Z}_2,
\eeq
where ${\bf Z}_2$ represents an intersection between $SU(2)_C$ and $U(1)_{{\rm
I}+{\rm Y}-2{\rm C}}$. Thus, in conclusion the manifold of rigidly gauge
equivalent fundamental monopoles is 
\beq
M(\calM_Q) = \frac{S(U(3)\x U(2))}
{SU(2)_C\x U(1)_{{\rm Y}-{\rm I}} \x U(1)_{{\rm I}+{\rm
Y}-2{\rm C}}/{\bf Z}_2}. 
\eeq
This is the first main result of this letter.

We should comment that it is possible to show all of the fundamental
monopoles lie within the same equivalence class. This is by associating the
different monopole embeddings with the spectrum of roots corresponding to the
roots of $SU(5)$ that are not roots of $S(U(3)\x U(2))$. The action of
$S(U(3)\x U(2))$ upon the associated root spaces takes one monopole
embedding to another. We shall discuss this fully in another
publication~\cite{me}. 

Now we shall consider the corresponding action of $S(U(3)\x U(2))$  upon a
fermion in the fundamental representations of colour, weak isospin and weak
hypercharge. In the standard model this corresponds to the $(u, d)_L$ quark
doublet. For $f_{(u,d)_L} \in {\bf C}^{3\x 2}$  the action is
\beq 
\label{fermion-gauge}
f_{(u,d)_L} \mapsto h_{\rm Y} h_{\rm C} \cdot \ f_{(u,d)_L}\cdot  h_{\rm I},
\eeq  
with $h_{\rm Y}$ interpreted as a complex phase and $h_{\rm C}$  and  
$h_{\rm I}$ elements  of $SU(3)_C$ and $SU(2)_I$ respectively. 

Consequently we may form a manifold of gauge equivalent fermion states
from the actions of $S(U(3)\x U(2))$ on this fermion $f_{(u,d)_L}$. The
manifold is of the form   
\beq
M(f_{(u,d)_L}) \cong \frac{S(U(3)\x U(2))}{C(f_{(u,d)_L})},
\eeq
with the stability group
\bea
C(f)= \{ h \in S(U(3)\x U(2))\ : h\cdot f=f\}
\eea
representing those transformations that leave $f$ invariant.

Without loss of generality we shall consider acting on the specific element
$f_{jk}=\de_{j1}\de_{k1}$. Again the generators used are normalised to the
topology of $S(U(3) \x U(2))$, 
\bea
C &=& i\: {\rm diag} (1, 1, -2),\\
I_3 &=& i\: {\rm diag} (1, -1),\\ 
Y &=& i\: {\rm diag} (1, 1),
\eea
such that $\exp (2 \pi Y) f = \exp (2 \pi I_3) f = \exp (2 \pi C) f = 1$.
In particular
\beq
\exp \theta_1 Y \exp \theta_3 C \cdot f \cdot \exp \theta_2 I_3 = 
e^{i(\theta_1+\theta_2+\theta_3)}f.
\eeq
From this we obtain
\beq
C(f_{(u,d)_L}) = SU(2)_C\x U(1)_{{\rm Y}-{\rm I}} \x U(1)_{{\rm Y}-2{\rm
C}}/{\bf Z}_2 
\eeq
and thus we conclude that the manifold of rigidly gauge equivalent fundamental
fermions is 
\beq
\label{fermionorbit}
M(f_{(u,d)_L}) = \frac{S(U(3)\x U(2))}{SU(2)_C\x U(1)_{{\rm Y}-{\rm I}} \x
U(1)_{{\rm Y}-2{\rm C}}/{\bf Z}_2}.
\eeq

By comparing the above manifolds we see that both are precisely the same
\beq
M(\calM_Q)=M(f_{(u,d)_L}).
\eeq
This is our main result. It shows an equivalence between the transformation
properties of fundamental monopoles and $(u,d)_L$ fermions. 
This supports that
fundamental monopoles transform under the same representation as the $(u,d)_L$
fermion. Namely the fundamental representation of $S(U(3)\x U(2))$.

We now consider the action of the dual group $S(U(3)\x U(2))^v=SU(3)\x SU(2)\x
U(1)$ on the fermion $f$. Then the associated gauge orbit is $S(U(3)\x
U(2))^v/C^v(f)$. However it is clear that $C^v(f)=C(f)\x {\bf Z}_6$. Thus
fermion gauge orbit in Eq.~(\ref{fermionorbit}) under $S(U(3)\x U(2))$ is the
same as the fermion gauge orbit under the dual group $S(U(3)\x U(2))^v$.
In other words the gauge orbits of monopoles are consistent with both the
residual symmetry group and the dual residual symmetry group.

It is an interesting feature of the above arguments that they imply an
association 
between the long range interactions of these monopoles and the gauge
interactions of a particle transforming under the fundamental representation of
$SU(3)_{\rm C}$, $SU(2)_{\rm I}$ and $U(1)_{\rm Y}$ gauge fields. In
particular note the transformations of Eq.~(\ref{mon-gauge}) are local. Then
the monopole moves around its gauge orbit under transformations of a local
symmetry. This feature should be viewed as the background to our work
on unification in the dual standard model~\cite{unif}. There the starting
assumption is that the monopoles interact via a gauge interaction, and as a
consequence we derive relations between the gauge couplings at monopole 
unification. 

Also of note is that the techniques used here relate purely to the symmetry
properties of the model. Thus our derivation should be very general, and we
expect that techniques used here will be applicable to other
situations of interest. Other examples that should be amenable to this
approach include monopoles from various symmetry breakings, and the long range
interactions of vortices.   

We now move on to a discussion of the gauge equivalence classes for the other
monopoles. These are formed from stable bound states of fundamental
monopoles~\cite{gard84}. 

Writing the magnetic charges of these other stable monopoles as
\beq
Q_{q_{\rm Y}}=\frac{1}{g_u} \left( q_{\rm C}T_{\rm C} + q_{\rm I}T_{\rm I} + 
q_{\rm Y}T_{\rm Y} \right), 
\eeq
where a particular state is labelled by its hypercharge,
the following spectrum of stable monopoles is obtained:
\begin{center}
\begin{tabular}{|c|ccc|ccc|c|}
\hline
\  & $q_{\rm C}$ & $q_{\rm I}$ & $q_{\rm Y}$ & $d_{\rm C}$ &
$d_{\rm I}$ & $d_{\rm Y}$ & f\\ 
\hline
$(e^{2i\pi/3}, -1)$ & 1 & 1/2 & 1/3 & 3 & 2 & 1& $(u,d)_L$ \\ 
$(e^{-2i\pi/3}, 1)$ & -1 & 0 & 2/3 & 3 & 0 & 1 & $\bar{d}_L$ \\
$(1, -1)$ & 0 & 1/2 & 1 & 0 & 2 & 1 & $(\bar{\nu},\bar{e})_R$\\
$(e^{2i\pi/3}, 1)$ & 1 & 0 & 4/3 & 3 & 0 & 1 & $u_R$\\
$(e^{-2i\pi/3}, -1)$ & - & - & - & - & - & - & - \\
$(1, 1)$ & 0 & 0 & 2 & 0 & 0 & 1 & $\bar{e}_L$\\
\hline
\end{tabular}
\end{center}
The degeneracies of each bound state has also been included, relating to the
degeneracy in Eqs.~(\ref{deg1},\ref{deg2},\ref{deg3}) and
Eq.~(\ref{deg4}). We have also included the standard model fermions
that have the same charges as the monopoles. 

We shall consider firstly the gauge equivalence classes of the fermions in the
standard model. As before these are of the form
\beq
M(f) = \frac{S(U(3)\x U(2))}{C(f)}
\eeq
with $C(f)$ the centraliser of the fermion's gauge transformations, namely
\beq
C(f) = \{ h \in S(U(3)\x U(2)): h\cdot f = f\}.
\eeq
A calculation analogous to that carried out in the first part of this letter
gives:
\bea
C(f_{(u,d)_L}) 
&=& SU(2)_{\rm C}\x U(1)_{{\rm Y}-{\rm I}} \x U(1)_{{\rm Y}-2{\rm C}}/{\bf
Z}_2\\ 
C(f_{\bar{d}_L}) 
&=& SU(2)_{\rm C}\x SU(2)_{\rm I} \x U(1)_{{\rm Y}-2{\rm C}}/{\bf Z}_2\\
C(f_{(\bar{\nu},\bar{e})_L})
&=& SU(3)_{\rm C}\x U(1)_{{\rm Y}-{\rm I}}\\
C(f_{u_R}) 
&=& SU(2)_{\rm C}\x SU(2)_{\rm I} \x U(1)_{{\rm Y}-2{\rm C}}/{\bf Z}_2\\
C(f_{\bar{e}_L})
&=& SU(3)_{\rm C}\x SU(2)_{{\rm I}}.
\eea

We shall now turn to the problem of determining the gauge
equivalence classes of the monopoles. However, our analysis is complicated by
the fact that higher charged stable monopoles
are not embedded monopoles. This point was crucial for our analysis in the
first part of this letter, where we associated a $\calM_Q$ with the monopole
embedding and described the group actions upon this.

Instead we shall deal only with the magnetic charge of the 
monopoles. Observe that for the $(u,d)_L$ fundamental monopole the subgroup of
$S(U(3)\x U(2))$ that leaves the magnetic charge invariant is
\beq
\!C(Q_{1/3})\! =\!\{h \in S(U(3)\!\x\! U(2))\!:
\Ad(h)Q_{1/3}\!=\!Q_{1/3}\}
\eeq
for which explicit calculation yields
\beq
C(Q_{1/3})
= SU(2)_{\rm C} \x U(1)_{\rm C} \x U(1)_{\rm I} \x U(1)_{\rm Y}/{\bf Z}_6.
\eeq
It is clear that this is related to $C(\calM_Q)$ by
\beq
C(Q_{1/3}) = C(\calM_Q) \x U(1)_Q/{\bf Z}_6.
\eeq
Physically this represents $U(1)_Q$ acting trivially upon $Q$, whilst acting
non-trivially upon the monopole. Whilst $U(1)_Q$ does not appear in the action
of $S(U(3)\x U(2))$ on the magnetic charge, it is still important in its
action upon the monopole embedding.

Now we verified in the first part of this letter that $C(\calM_Q) =
C(f_{(u,d)_L})$. In fact this was all that was needed to prove that the 
monopole and fermion gauge equivalence classes were the same. Taking the
analogy of this we shall show that 
\beq
\label{rel}
C(f_{(u,d)_L})\x U(1)_Q/{\bf Z}_6 = C(Q),
\eeq
for each of the respective higher charge monopoles and their associated
fermions. 

The magnetic charges of the higher charge monopoles are given by the table 
above. From this we calculate
\bea
Q_{1/3}^{r+}&=& i\:{\rm diag}(1,0,0,0,-1),\\
Q_{2/3}^r&=& i\:{\rm diag}(0,1,1,-1,-1),\\
Q_{1}^+&=& i\:{\rm diag}(1,1,1,-1,-2),\\
Q_{4/3}^r&=& i\:{\rm diag}(2,1,1,-2,-2),\\
Q_{2}&=& i\:{\rm diag}(2,2,2,-3,-3).
\eea
which yields their respective stability groups 
\bea
C(Q_{1/3})
&=& U(2)_{\rm C} \x U(1)_{\rm I} \x U(1)_{\rm Y}/{\bf Z}_6\\
C(Q_{2/3})
&=& U(2)_{\rm C} \x SU(2)_{\rm I} \x 
U(1)_{\rm Y}/{\bf Z}_6\\
C(Q_{1})
&=& SU(3)_{\rm C} \x U(1)_{\rm I} \x U(1)_{\rm Y}/{\bf Z}_6\\
C(Q_{4/3})
&=& U(2)_{\rm C} \x SU(2)_{\rm I} 
\x U(1)_{\rm Y}/{\bf Z}_6\\
C(Q_{2})
&=& 
SU(3)_{\rm C} \x SU(2)_{\rm I} \x U(1)_{\rm Y}/{\bf Z}_6.
\eea
From this it is a simple matter to see that Eq.~(\ref{rel}) holds.

However, the above does not rigourously prove equivalence of their gauge 
equivalence classes; to do that one must examine the specific  
form of the monopoles, as in the first part of this
letter. Nevertheless the verification that Eq.~(\ref{rel}) holds for each of
the monopoles and their respective fermion counterparts constitutes a
strong indication that the gauge equivalence classes are the same.

We conclude this letter with a last remark about the structure of the
higher charge monopole equivalence classes. Consideration of the above
equations reveals that the $(\nu,e)_L$ monopole does not
transform under colour symmetry. Thus it is naturally associated with
fundamental monopoles arising from the symmetry breaking
\beq
SU(3) \rightarrow U(2)= SU(2)_{\rm I}\x U(1)_{\rm Y}/{\bf Z}_2.
\eeq
These monopoles are again given by embedding an $SU(2)$ monopole. Then the
gauge equivalence class of such fundamental monopoles are determined by
analogous methods to those in the first part of this letter
\beq
M(\calM_{Q_1}) \cong \frac{U(2)}{U(1)_{Y-I}}.
\eeq
This is the same manifold as the gauge equivalence class of $(\nu,e)_L$
fermions. 

Likewise the monopoles associated with $u_R$ and $d_R$ do not transform under
isospin symmetry and it is natural to associate them with monopoles arising
from
\beq
SU(4) \rightarrow U(3)= SU(3)_{\rm C} \x U(1)_Y/{\bf Z}_3.
\eeq
Here the $d_R$ monopoles is given by embedding an $SU(2)$ monopole, whilst the
$u_R$ is interpreted as a bound state of two of these. Their gauge equivalence
class is
\beq
M(\calM_{Q_{2/3}}) \cong \frac{U(3)}{SU(2)_C \x U(1)_{Y-2C}/{\bf Z}_2},
\eeq
the same as for the $u_R$ fermion.

Finally the monopole associated with $e_R$ is associated with monopoles
arising from
\beq
SU(2) \rightarrow U(1)_Y.
\eeq
Again, trivially, this is an embedded monopole. This time the gauge
equivalence class is 
\beq
M(\calM_{Q_2}) \cong U(1),
\eeq
the same as for $e_R$.

Thus we remark that the higher charge monopoles are associated with
fundamental monopoles in other symmetry breakings. Furthermore their gauge
equivalence classes are calculable by similar methods to the first part of
this letter. Such calculations yield the same equivalence classes as the
corresponding fermions. 


{\it
I acknowledge King's College, Cambridge for a junior research
fellowship and P.~Saffin for discussions. 
}


\end{document}